\documentclass[preprint]{aastex}

\begin{document}

\title{The Stellar Content of NGC~6789, A Blue Compact
Dwarf Galaxy in the Local Void\footnotemark[1] }

\author{Igor O. Drozdovsky}
\affil{University of Pittsburgh, Pittsburgh, PA 15260, USA, and
Astronomical Institute, St.Petersburg State University, Russia}
\email{dio@phyast.pitt.edu}
\author{Regina E. Schulte-Ladbeck}
\affil{University of Pittsburgh, Pittsburgh, PA 15260, USA}
\email{rsl@phyast.pitt.edu}
\author{Ulrich Hopp}
\affil{Universit\"{a}tssternwarte M\"{u}nchen, M\"{u}nchen, FRG} 
\email{hopp@usm.uni-muenchen.de}
\author{Mary M. Crone}
\affil{Skidmore College, Saratoga Springs, NY 12866, USA}
\email{mcrone@skidmore.edu}
\author{Laura Greggio}
\affil{Osservatorio Astronomico di Bologna, Bologna, Italy, and
Universit\"{a}tssternwarte M\"{u}nchen, M\"{u}nchen, FRG}
\email{greggio@usm.uni-muenchen.de}

\footnotetext[1]{Based on observations made with the NASA/ESA Hubble 
Space Telescope obtained from the 
Space Telescope Science Institute, which is operated by the Association of 
Universities for Research in Astronomy, Inc., under NASA contract NAS
5-26555.}

\date{\today}

\begin{abstract}
We find that NGC~6789 is the most nearby example of a
Blue Compact Dwarf galaxy known to date. With the help of WFPC2 aboard the 
Hubble Space Telescope,
we resolve NGC~6789 into over 15,000 point sources in the V and I bands. 
The young stars of NGC~6789 are found exclusively near the center of the
galaxy.
The red giant population identified at large galacticentric radii yields 
a distance of about 3.6~Mpc, a stellar metallicity [Fe/H] of 
about -2, and a minimum age of about 1~Gyr. Despite its isolated location
in the Local Void,
its low metallicity, and
its active star formation, the properties of NGC~6789 are clearly not those of
a galaxy in formation. 

\end{abstract}


\section{Introduction}

Galaxies in voids are interesting for their potential use as
probes of Cold Dark Matter (CDM) cosmogonies (Dekel \& Silk 1986).
In general, these models predict that isolated galaxies are less
evolved than those in denser environments.  For example, N-body
simulations show that dark matter halos in lower-density regions
 tend to have lower masses (e.g. Lemson \& Kauffmann 1999), making
them more susceptible to negative feedback from ionizing radiation
and stellar winds (Ferrara \& Tolstoy 2000).  Indeed, recent high-resolution 
simulations which incorporate
a semi-analytic prescription for star formation specifically predict 
that star formation is inhibited in voids
(White \& Springel 2000). Clearly, these models should be 
tested against the observed star-formation history of void galaxies.

The most straightforward way to observe the star-formation history
of galaxies is through deep single-star photometry, which can directly
identify stars in various evolutionary phases through their positions
on a color-magnitude diagram (CMD).
While observations
of chemical abundances can be performed for more distant star-forming
galaxies, 
the validity of age-dating from certain element abundance
ratios has been challenged by recent chemical evolution models, which indicate
that even extremely low-metallicity galaxies can have formed many Gyr ago 
(Henry, Edmunds \& K\"{o}ppen 2000).   

When Karachentsev \& Makarov (1998) presented NGC~6789 as an unevolved dwarf
galaxy (akin to NGC~1705 and NGC~2915, e.g. Meurer et al. 1992) in the
direction of the Local Void in Hercules-Aquila
(with a heliocentric velocity of -147~km~s$^{-1}$), 
we had found the perfect candidate for a stellar 
population study of a low-mass void galaxy with the HST.
NGC~6789 is extremely isolated (according to a NED query
with a $40\degr$ search radius and Karachentsev, private communication);
it is separated by at least 2.5~Mpc from the massive galaxy NGC~6946.
NGC~6789 has the properties of Blue Compact Dwarfs (BCDs),
systems long considered candidates for young
galaxies in the nearby Universe (Sargent \& Searle 1970). 

Meanwhile, Drozdovsky \& Tikhonov (2000, hereafter DT) imaged NGC~6789 from
the ground.
They found that it belongs to the iE subtype, which exhibits the morphology
most characteristic of BCDs. This result, together with our work
on the distribution of red giants in the iE BCD VII~Zw~403 (Schulte-Ladbeck
et al.
1999), already suggests
the presence of stellar generations which pre-date the on-going starburst. DT
identified the faint red stars in their CMD with
the tip of the red giant branch (TRGB),
and derived a very short distance of only 2.1~Mpc (m-M=26.6). However,
as our HST data show, DT identified bright asymptotic giant branch (AGB)
stars with the TRGB, severely underestimating the true distance of NGC~6789. 
Nevertheless, in this paper we show that NGC~6789 is still the most nearby
BCD on record;
and the presence of RGB stars allows us to ascertain that it is at least
about 1~Gyr old.

\section{Observations and reductions}

Observations of NGC~6789 were obtained July through September of 2000 with
the WFPC2 aboard HST
as part of GO programs 8645 and 8122. In Fig.~1, we
overlay deep WFPC2 photometry on the Nordic Optical Telescope (NOT) image
of DT. The central star-forming region was encompassed
by the PC chip, whereas the WF chips were pointed into the
low-surface-brightness
sheet. Resolved halo stars can be traced to large distances from the center
of NGC~6789
(see the physical scale of Fig.~1).

We discuss photometry based on dithered exposures in the F555W (V) and
F814W (I) bands taken
as part of our program 8122. We performed the usual post-pipeline
reductions and used 
drizzling to combine individual exposures. The total exposure times are 
8,200s each in V and I. Fig.~2 shows the color image of the PC, exhibiting
many resolved stars. We performed single-star photometry 
with DAOPHOT~II (Stetson 1994). The WF chips have a low stellar
density ($\sim$~0.4~stars/\sq\arcsec;
showing many background galaxies which were rejected in the photometry); and
we could easily find isolated stars to model the
point-spread function (PSF). On the contrary, the
PC chip has a very high stellar density ($\sim$~6.7~stars/\sq\arcsec), much
higher than what we encountered in our analysis of UGCA 290
($\sim$~2.0~stars/\sq\arcsec; Crone et al. 2000) or
VII~Zw~403 ($\sim$~2.8~stars/\sq\arcsec; Schulte-Ladbeck et al. 1999).
It was difficult to find suitable stars to model the PSF, and these data are
clearly affected by blending due to crowding in a substantial way. This leads
to a shallow detection limit for the PC data (see Fig.~3a). We took 
zero points from the May 1997 SYNPHOT tables, and corrected for foreground
extinction
in the direction of NGC~6789 using the values of Schlegel, Finkbeiner \&
Davis (1998)
with R$_V$ = 3.1: A$_V$ = 0.232 and A$_I$ = 0.136. We did not correct for
internal extinction within NGC~6789, but the transparency of the halo
population
(the population from which we derive the distance) is high and suggests that
the extinction is
low at high galacticentric radii. We performed ADDSTAR tests on the WF chips
and found that completeness is high ($>$90\%) well below the TRGB.
 
\section{Results}

The CMDs of NGC~6789 illustrate the spatial
variation
in its stellar content from the ``core" to the ``halo". The CMD of the PC
(Fig.~3a)
exhibits a stellar mix which is characteristic of the superposition
of many stellar generations. The ``blue plume" at (V-I) $\sim$ 0 contains 
massive blue supergiants
and main-sequence stars, the ``red plume" at (V-I) $\sim$~1.5 contains
evolved supergiants
and AGB stars, the ``red tail" extending past (V-I) $>$~2 contains
intermediate-mass AGB stars in the thermally pulsing phase, and the
concentration
of stars below the TRGB contains low-mass red giants and AGB stars. For
reference,
we overlay the CMD with the Z=0.0004 and 0.004 Padova stellar evolutionary
tracks
(Fagotto et al. 1994) assuming the distance derived below. 
The blue plume is quite wide, and the TRGB looks slightly brighter than
tip of 1~M$_\sun$ tracks. Artificial star tests show that crowding and
blending dominate these effects, while unknown internal extinction
variations may also contribute.

The WF chips lack blue stars down to the detection limit (see Fig.~3b).
This suggests that
the average population age is higher in the halo than it is in the core. 
The CMD of the halo is characterized by
a comparatively narrow red giant branch. We overplot onto the CMD a few
Globular Cluster ridgelines for low-metallicity clusters from
Da Costa \& Armandroff (1990).
Additional stars scattered across the CMD are attributed
to galactic foreground contamination: the galactic latitude of NGC~6789
is only 21$\fdg$5
(and see Fig.~1).

We used the WF data to find the TRGB, and to determine the distance and
metallicity of
NGC~6789 (following Lee, Freedman \& Madore 1993). The WF data are preferred
over the PC data since they exhibit much less crowding, and fewer AGB stars.
There are sufficiently large numbers of stars found on
each chip, that we were able to construct I-band luminosity functions
for each chip individually. The location of the TRGB occurs at the same
magnitude in
all three chips. The final TRGB magnitude was derived from the combined
I-band luminosity
function of about 7,000 stars selected to have 0.7$<$(V-I)$_0$$<$2.0 (see
Fig.~4). The TRGB was found by applying a Sobel filter; it occurs at
I$_0$ = 23.82$\pm$0.05. In order to derive the metallicity of the RGB stars
we measure the (V-I)$_0$ color both at half a
magnitude below the
TRGB and, owing to the high quality of the data, also at one magnitude
below the TRGB.
We consistently find a low metallicity, which, at 0.5 mag below the TRGB, 
is [Fe/H] = $-1.92\pm0.03\pm0.15$,
where the first error is the uncertainty in the mean and the second is
the systematic error.
With the appropriate bolometric correction, the absolute magnitude of the
TRGB is -4.0, yielding a distance modulus of $27\fm80\pm0\fm13\pm0\fm18$,
or a distance of $3.6\pm0.2\pm0.3$~Mpc.
This location of the TRGB is indicated in Fig.~3a and 3b by a dashed line.

NGC~6789 is about 70\% more distant than was derived
by DT. At this new distance, an angular scale of 1$\arcmin$ corresponds to
a physical
scale of 1.05~kpc. The star-formation rate
derived by DT from the H$_{\alpha}$ luminosity increases to
0.04~M$_{\odot}$yr$^{-1}$
(for a Salpeter initial mass function ranging from 0.1--100~M$_{\odot}$, i.e.,
Hunter \& Gallagher 1986). 
The integrated total magnitudes of NGC~6789 within the
$\mu_{\rm V}=25~{\rm mag/\sq\arcsec}$ isophote of DT,
$m_{\rm I}\!=\!13\fm63\pm0\fm15$ and $m_{\rm V}\!=\!14\fm62\pm0\fm15$,
correspond to absolute magnitudes of
$M_{\rm I,0}=-14\fm31\pm~0\fm20$ and $M_{\rm V,0}=-13\fm41\pm~0\fm20$.
DT show that the outer regions of NGC~6789 exhibit the exponential surface 
brightness profile of a disk galaxy.
The scale length correponds to 280~pc, much smaller than that of VII~Zw~403
($\approx$600~pc, Schulte-Ladbeck et al. 1999).
S.~Moehler kindly obtained a spectrum of NGC~6789 for us from which we
measured the strengths of [O{\sc ii}], [O{\sc iii}], [N{\sc ii}] and
H-Balmer emission lines. The temperature-sensitive [O{\sc iii}] line at
4636\AA \phantom{-} is undetected. This leaves 12 + log(O/H) two-valued
(Edmunds \& Pagel 1984) at 7.7 or 8.5.
While NGC 6789 appears to have a sub-solar oxygen abundance,
its N/O ratio (estimated following Thurston, Edmunds \& Henry 1996) is
larger than the minimum value seen in BCDs, log(N/O) = -1.6 (Izotov \&
Thuan 1999). Therefore, NGC~6789 is probably not one of the rare extremely
metal-poor BCDs resembling I~Zw~18, but rather, a BCD with moderately low
ionized gas metallicity.

NGC~6789 is undetected in
H{\sc i} (Huchtmeier et al. 2000). Possible reasons given are confusion with
local H{\sc i} or lack of sensitivity for very weak emission.
If NGC~6789 could indeed be shown to possess very little neutral gas,
it could be interpreted as a BCD near the end of its life.

\section{Discussion}

We start by comparing the properties
of NGC~6789 to those of the other four BCDs which are
resolved with HST to magnitudes below the TRGB.
NGC~6789, at a mere 3.6~Mpc, is the most nearby BCD resolved
to date. Ordered by increasing distance we have: 
Mrk~178 ($\geq$4.2$\pm$0.5~Mpc, Schulte-Ladbeck et al. 2000),
VII~Zw~403 (4.4$\pm$0.2~Mpc, Schulte-Ladbeck et al. 1999), I~Zw~36
($\geq$5.8$\pm$0.4~Mpc, Schulte-Ladbeck
et al. 2001), and UGCA~290 (6.7$\pm$0.4~Mpc, Crone et al. 2000).
NGC~6789, situated in the Local Void, is the BCD in the most underdense
region. 
VII~Zw~403 is on the line of sight to M~81 but
at least a Mpc behind the center of the group.
Mrk~178, I~Zw~36 and UGCA~290 all lie in the direction of
the Canes Venatici Cloud.

The stellar population in NGC~6789 exhibits several features typical
of the other resolved BCDs. 
For example, the CMD of the star-forming core of NGC~6789
exhibits a prominent population of luminous AGB stars. 
This serves to indicate that star formation
in the centers of BCDs (at the bottom of the gravitational well)
has continued to be active over the past several hundred Myr.
All of the nearby BCDs resolve into red giants, which in all
but UGCA~290, are much more extended spatially than the young, main sequence
and AGB stars. We have termed
these halos ``Baade's red sheets" (Schulte-Ladbeck, Crone \& Hopp 1998).
Because stars of a wide range of ages overlap on the RGB, we can only
be sure that these stars are at least 1~Gyr old, while they could in principle
be as old as the oldest stars in the Universe.
Whenever optical photometry is available we find that the 
halo populations of BCDs exhibit blue RGBs, which can be interpreted
to indicate low mean stellar [Fe/H] ratios of about -2 (for VII~Zw~403,
-1.92$\pm$0.04,
for UGCA~290, -2.0$\pm$0.1), as low as those of the
most metal poor galactic Globular Clusters.

The current star-formation rate of NGC~6789 is not particularly high.
The rates are never exceptionally high in the cases we have investigated, only
of order a few times
10$^{-3}$--10$^{-2}$M$_{\odot}$yr$^{-1}$, from either or both, H$_{\alpha}$
luminosity
and CMD modeling. They are orders of magnitude below the high rates, 
10${^0}$--10${^2} $M$_{\odot}$yr$^{-1}$, determined for a sample of 27 BCDs
in the Second
Byurakan Survey (Izotova, Parnovsky \& Izotov 2000).

The question of whether BCDs in voids are statistically different from
those in the field is an area of active research.  V\'{\i}lchez (1995)
compared BCDs in the Virgo cluster with isolated BCDs.
He found that BCDs in low density environments are likely to exhibit their
first
major episode of star formation.  More recently, however, Popescu, Hopp \&
Rosa (1999) investigated a
sample of BCDs in voids and in the field, and found no differences between
their
star-formation rates or metallicities.  Moreover, Vennik, Hopp, \& Popescu
(2000) performed B and R surface photometry on the
above galaxy sample, and found that field and void BCDs are indistinguishable
in terms of luminosity and color. They also noted that more than half of
their void BCDs show indications for low surface-brightness
red disks, possibly due to an older stellar population.

Our results for the most nearby BCDs support the idea that void and
field BCDs are similar, in terms of resolved stellar content,
star-formation rate,  Fe/H ratio, and morphology.  In particular, they
support the hypothesis that void BCDs have stellar background sheets that
pre-date the on-going starburst.

The ubiquity of Baade's sheet in {\it every} galaxy resolved
deeply enough to detect red giants~--- including isolated and
low-metallicity star-forming dwarfs~--- imposes significant constraints
on models of galaxy formation and chemical evolution.

\acknowledgments Work on this project was supported 
through HST grants to RSL and MMC (project 8122). 
UH acknowledges financial support from SFB~375. 
LG acknowledges support from
the Alexander von Humboldt Stiftung.
We thank S.~Moehler for providing us with her NGC~6789 spectrum.
We made extensive use of the SIMBAD and NED data bases.

\clearpage

\figcaption[] {A color picture of the WFPC2-resolved stars from program 8122
overlayed onto the NOT
ground-based image of DT. N is up and E is to the left. The size of the image
is $3\farcm7\times3\farcm7$, which corresponds to $\sim 4\times4$~kpc.
Notice how far out the resolved stars can be traced.
}

\figcaption[] {A color image of the PC chip, illustrating the compact, dense
core of star formation. The color combinations used are: red = F814W,
green = F555W+F814W+F656N, and blue = F555W+F336N.
}

\figcaption[] { a) The CMD of the PC chip with a range of stellar tracks
overlayed for two metallicities. Tracks with metallicities Z=0.004 and
0.0004 are marked by red and green colors respectively.
b) The CMD of the WF chips with low Fe/H Globular Cluster ridge lines
overlayed. The dashed line on both images indicates the position of the TRGB.
}

\figcaption[] {The I-band luminosity function (solid line), and the
Sobel-filtered
luminosity function (dotted line). The location of the TRGB is indicated.
}

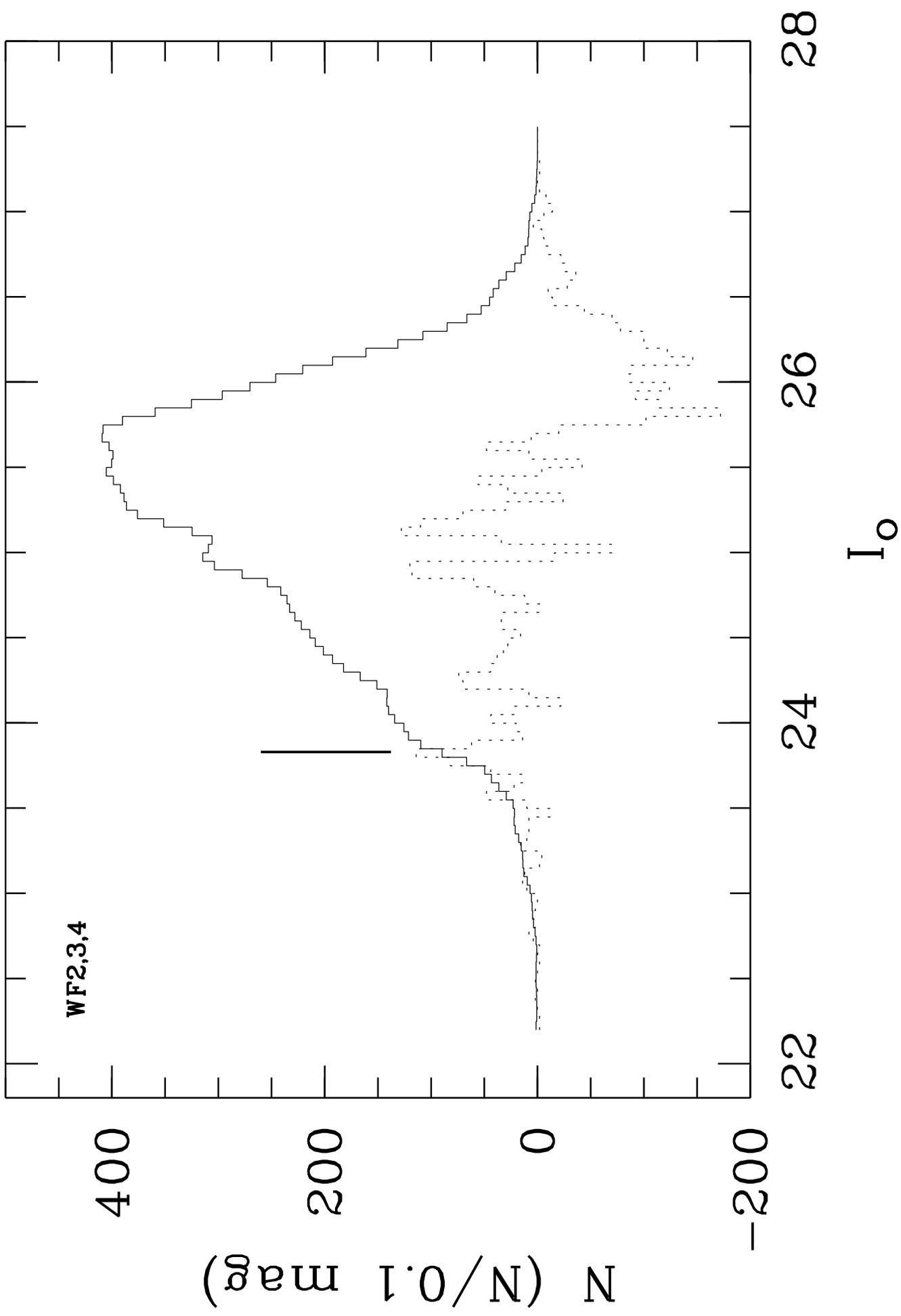
\end{document}